\documentclass[manuscript]{aastex63}

\usepackage{hyperref}

\hypersetup{pdftitle= {Preliminary data}
pdfauthor = {SNf},
colorlinks=true,        
linkcolor=blue,         
citecolor=blue,         
urlcolor=blue           
}
 
\begin{document}
\title{The SNEMO and SUGAR Companion Datasets}

\author{     G.~Aldering}
\affiliation{    Physics Division, Lawrence Berkeley National Laboratory,
    1 Cyclotron Road, Berkeley, CA, 94720}

\author{     P.~Antilogus}
\affiliation{ Laboratoire de Physique Nucl\'eaire et de Hautes \'Energies, CNRS/IN2P3, Sorbonne Universit\'e, Paris Diderot, F-75005, Paris, France}

\author{    C.~Aragon}
\affiliation{ Physics Division, Lawrence Berkeley National Laboratory,
    1 Cyclotron Road, Berkeley, CA, 94720}

\author{     S.~Bailey}
\affiliation{    Physics Division, Lawrence Berkeley National Laboratory,
    1 Cyclotron Road, Berkeley, CA, 94720}

\author{     C.~Baltay}
\affiliation{    Department of Physics, Yale University,
    New Haven, CT, 06250-8121}

\author{     S.~Bongard}
\affiliation{ Laboratoire de Physique Nucl\'eaire et de Hautes \'Energies, CNRS/IN2P3, Sorbonne Universit\'e, Paris Diderot, F-75005, Paris, France}

\author{     K.~Boone}
\affiliation{    Physics Division, Lawrence Berkeley National Laboratory,
    1 Cyclotron Road, Berkeley, CA, 94720}
\affiliation{
    Department of Physics, University of California Berkeley,
    366 LeConte Hall MC 7300, Berkeley, CA, 94720-7300}

\author{     C.~Buton}
\affiliation{    Universit\'e de Lyon, F-69622, Lyon, France ; Universit\'e de Lyon 1, Villeurbanne ;
    CNRS/IN2P3, Institut de Physique Nucl\'eaire de Lyon}

\author{     N.~Chotard}
\affiliation{    Universit\'e de Lyon, F-69622, Lyon, France ; Universit\'e de Lyon 1, Villeurbanne ;
    CNRS/IN2P3, Institut de Physique Nucl\'eaire de Lyon}

\author{     Y.~Copin}
\affiliation{    Universit\'e de Lyon, F-69622, Lyon, France ; Universit\'e de Lyon 1, Villeurbanne ;
    CNRS/IN2P3, Institut de Physique Nucl\'eaire de Lyon}

\author{     S.~Dixon}
\affiliation{    Physics Division, Lawrence Berkeley National Laboratory,
    1 Cyclotron Road, Berkeley, CA, 94720}
\affiliation{
    Department of Physics, University of California Berkeley,
    366 LeConte Hall MC 7300, Berkeley, CA, 94720-7300}

\author{     H.~K.~Fakhouri}
\affiliation{    Physics Division, Lawrence Berkeley National Laboratory,
    1 Cyclotron Road, Berkeley, CA, 94720}
  \affiliation{
    Department of Physics, University of California Berkeley,
    366 LeConte Hall MC 7300, Berkeley, CA, 94720-7300}

\author{     U.~Feindt}
\affiliation{The Oskar Klein Centre, Department of Physics, AlbaNova, Stockholm University, SE-106 91 Stockholm, Sweden}

\author{     D.~Fouchez}
\affiliation{    Centre de Physique des Particules de Marseille,
    Aix-Marseille Universit\'e , CNRS/IN2P3,
    163 avenue de Luminy - Case 902 - 13288 Marseille Cedex 09, France}

\author{     E.~Gangler}
\affiliation{    Universit\'e Clermont Auvergne, CNRS/IN2P3, Laboratoire de Physique de Clermont, F-63000 Clermont-Ferrand, France}

\author{     B.~Hayden}
\affiliation{    Physics Division, Lawrence Berkeley National Laboratory,
    1 Cyclotron Road, Berkeley, CA, 94720}

\author{     W.~Hillebrandt}
\affiliation{    Max-Planck-Institut f\"ur Astrophysik, Karl-Schwarzschild-Str. 1,
D-85748 Garching, Germany}

\author{A.~G.~Kim}
\affiliation{    Physics Division, Lawrence Berkeley National Laboratory,
    1 Cyclotron Road, Berkeley, CA, 94720}

\author{     M.~Kowalski}
\affiliation{    Institut fur Physik,  Humboldt-Universitat zu Berlin,
    Newtonstr. 15, 12489 Berlin}
\affiliation{ DESY, D-15735 Zeuthen, Germany}

\author{     D.~K\"usters}
\affiliation{    Institut fur Physik,  Humboldt-Universitat zu Berlin,
    Newtonstr. 15, 12489 Berlin}

\author{     P.-F.~L\'eget}
\affiliation{ Laboratoire de Physique Nucl\'eaire et de Hautes \'Energies, CNRS/IN2P3, Sorbonne Universit\'e, Paris Diderot, F-75005, Paris, France}

\author{  Q.~Lin}
\affiliation{  Tsinghua Center for Astrophysics, Tsinghua University, Beijing 100084, China}

\author{     S.~Lombardo}
\affiliation{    Institut fur Physik,  Humboldt-Universitat zu Berlin,
    Newtonstr. 15, 12489 Berlin}

\author{ F.~Mondon}
\affiliation{    Universit\'e Clermont Auvergne, CNRS/IN2P3, Laboratoire de Physique de Clermont, F-63000 Clermont-Ferrand, France}

\author{     J.~Nordin}
\affiliation{    Institut fur Physik,  Humboldt-Universitat zu Berlin,
    Newtonstr. 15, 12489 Berlin}

\author{     R.~Pain}
\affiliation{ Laboratoire de Physique Nucl\'eaire et de Hautes \'Energies, CNRS/IN2P3, Sorbonne Universit\'e and Universit\'e de Paris, 4 Place Jussieu, 75005, Paris, France}

\author{     E.~Pecontal}
\affiliation{   Centre de Recherche Astronomique de Lyon, Universit\'e Lyon 1,
    9 Avenue Charles Andr\'e, 69561 Saint Genis Laval Cedex, France}

\author{    R.~Pereira}
 \affiliation{    Universit\'e de Lyon, F-69622, Lyon, France ; Universit\'e de Lyon 1, Villeurbanne ;
    CNRS/IN2P3, Institut de Physique Nucl\'eaire de Lyon}

 \author{    S.~Perlmutter}
 \affiliation{    Physics Division, Lawrence Berkeley National Laboratory,
    1 Cyclotron Road, Berkeley, CA, 94720}
\affiliation{
    Department of Physics, University of California Berkeley,
    366 LeConte Hall MC 7300, Berkeley, CA, 94720-7300}
    
\author{ K.~Ponder}
\affiliation{ Physics Division, Lawrence Berkeley National Laboratory, 1 Cyclotron Road, Berkeley, CA, 94720}
\affiliation{Berkeley Center for Cosmological Physics,
    University of California Berkeley,
    341 Campbell Hall, Berkeley, CA 94720, USA  }

\author{  M.~Pruzhinskaya}
\affiliation{ Universit\'e Clermont Auvergne, CNRS/IN2P3, Laboratoire de Physique de Clermont, F-63000 Clermont-Ferrand, France}
\affiliation{ Lomonosov Moscow State University, Sternberg Astronomical Institute, Universitetsky pr. 13, Moscow 119234, Russia}

 \author{    D.~Rabinowitz}
 \affiliation{    Department of Physics, Yale University,
    New Haven, CT, 06250-8121}

 \author{    M.~Rigault}
 \affiliation{    Universit\'e Clermont Auvergne, CNRS/IN2P3, Laboratoire de Physique de Clermont, F-63000 Clermont-Ferrand, France}

 \author{    D.~Rubin}
 \affiliation{  Department of Physics and Astronomy, University of Hawai?i at M?anoa, Honolulu, Hawai?i 96822}

 \author{    K.~Runge}
 \affiliation{    Physics Division, Lawrence Berkeley National Laboratory,
    1 Cyclotron Road, Berkeley, CA, 94720}

\author{    C.~Saunders}
\affiliation{ Princeton University, Department of Astrophysics, 4 Ivy Lane, Princeton, NJ, 08544}

\author{ L.-P.~Says}
\affiliation{ Universit\'e Clermont Auvergne, CNRS/IN2P3, Laboratoire de Physique de Clermont, F-63000 Clermont-Ferrand, France}

\author{     G.~Smadja}
\affiliation{    Universit\'e de Lyon, F-69622, Lyon, France ; Universit\'e de Lyon 1, Villeurbanne ;
    CNRS/IN2P3, Institut de Physique Nucl\'eaire de Lyon}

\author{    N.~Suzuki}
\affiliation{    Physics Division, Lawrence Berkeley National Laboratory,
    1 Cyclotron Road, Berkeley, CA, 94720}

\author{     C.~Tao}
\affiliation{   Tsinghua Center for Astrophysics, Tsinghua University, Beijing 100084, China }
\affiliation{    Centre de Physique des Particules de Marseille,
    Aix-Marseille Universit\'e , CNRS/IN2P3,
    163 avenue de Luminy - Case 902 - 13288 Marseille Cedex 09, France}

\author{     S.~Taubenberger}
\affiliation{    Max-Planck-Institut f\"ur Astrophysik, Karl-Schwarzschild-Str. 1,
D-85748 Garching, Germany}

\author{     R.~C.~Thomas}
\affiliation{    Computational Cosmology Center, Computational Research Division, Lawrence Berkeley National Laboratory,
    1 Cyclotron Road MS 50B-4206, Berkeley, CA, 94720}
 
\author{    M.~Vincenzi}
\affiliation{    Physics Division, Lawrence Berkeley National Laboratory,
    1 Cyclotron Road, Berkeley, CA, 94720}
    
\author{ B.~Weaver}
\affiliation{ National Optical-Infrared Astronomy Research Laboratory, 950 North Cherry Avenue, Tucson, AZ 85719, USA}

\collaboration{44}{The Nearby Supernova Factory}

\shortauthors{SNfactory}

\section*{}

The Nearby Supernova Factory (SNfactory, \citealt{Aldering:2002}) has made spectrophotometric observations of Type Ia supernovae since $2004$. This work presents an interim version of the data produced, including $210$ supernovae observed between $2004$ and $2013$. This data was taken with the SuperNova Integral Field Spectrograph (SNIFS, \citealt{Lantz:2004}), an instrument continuously mounted on the University of Hawaii $2.2$ m telescope on Maunakea, which is optimized for automated observation of point sources on a structured background over the full ground-based optical window at moderate spectral resolution. The supernovae have been flux-calibrated as described in \citet{Buton:2012} and \cite{Pereira:2013}, and the host-galaxy has been subtracted as described in \cite{Bongard:2011}. A publication containing the full description of the data pipeline is in preparation. The supernovae included in this release have been corrected for Milky Way dust extinction (\citealt{Schlegel:1998}, \citealt{Cardelli:1989}) and have been shifted to a common restframe at $z=0$. The absolute magnitudes were shifted by a random value to preserve blinding for ongoing cosmology analyses.

Two slightly different versions of the SNfactory data are presented here, which we label ``NewMexico-Caballo'' and ``NewYork-Alleg'' (short for Allegheny), following a naming convention in which a flux calibration run is given the name of a US state and the corresponding host-subtracted, extracted spectra are given the name of a park in that state. NewMexico-Caballo and NewYork-Alleg are respectively the datasets used in the training and validation of the SUGAR\footnote{supernovae.in2p3.fr/sugar} (\citealt{Leget:2019}) and SNEMO\footnote{snfactory.lbl.gov/snemo} (\citealt{Saunders:2018}) supernova models. The SNEMO training data is available at \href{cdsarc.u-strasbg.fr/viz-bin/cat/J/ApJ/869/167}{cdsarc.u-strasbg.fr/viz-bin/cat/J/ApJ/869/167} and the SUGAR training data is available at \href{cdsarc.u-strasbg.fr/viz-bin/cat/J/A+A/636/A46}{cdsarc.u-strasbg.fr/viz-bin/cat/J/A+A/636/A46}. These datasets were also used in additional SNfactory science analyses performed during the same time period (\citealt{Huang:2017}, \citealt{Nordin:2018}, \citealt{Leget:2018}, \citealt{Rigault:2018}). 

The SNEMO companion dataset, NewYork-Alleg, was produced slightly later than the SUGAR dataset, NewMexico-Caballo, and thus includes more recent supernovae and reflects changes related to the incremental improvements to the data pipeline. The analyses also used different criteria for their data selection with respect to the required phase coverage and inclusion of peculiar Type Ia supernovae, as described in \cite{Leget:2019} and \cite{Saunders:2018}. This results in different sets of supernovae in NewMexico-Caballo and NewYork-Alleg. The number of supernovae and spectra included in each dataset are described in Table~\ref{table: 1}. 

\begin{table*}
\begin{tabular}{llll}
\hline \hline 
Data properties & SNEMO companion dataset & SUGAR companion dataset & In Common\\ 
\hline 
Total number of SNe Ia & 171 &  172 & 133\\ 
Total number of spectra & 2485 & 2187  & 1702\\ 
Rest frame wavelength range & 3305.0 \AA \ -- 8586.0 \AA &  3273.6 \AA \ -- 8657.7 \AA \\ 
\hline 
\end{tabular} 
\caption{Numbers of supernovae and spectra in the SUGAR and SNEMO companion datasets} 
\label{table: 1} 
\end{table*} 


In addition to being blinded with respect to the absolute magnitude, the supernova data productions presented here are not calibrated to a level that would allow them to be used for cosmology. The variance spectrum included in NewMexico-Caballo and NewYork-Alleg only reports the scatter of the data due to photon and readout noise. The methodology in \cite{Leget:2019} and \cite{Saunders:2018} appropriately dealt with the total variance affecting their respective models. Additionally, the model construction methods in these analyses are not sensitive to the absolute calibration, since they focus on spectral differences in the supernovae, and are not sensitive to relative differences in the brightness of the supernovae, nor to their absolute magnitudes.

Recent and current work has resulted in a number of significant improvements in the SNfactory data with respect to its ties to the standard star network, calibration on `non-photometric' nights, and other aspects of the data processing. These improvements will enable the dark energy equation of state to be measured; a corresponding reprocessing of the SNfactory data and accompanying publications are in preparation.

\section*{Acknowledgments}
The authors are grateful to the technical and scientific staff of the University of Hawaii 2.2 m telescope, the Palomar Observatory, and the High Performance Wireless Radio Network (HPWREN). We thank Dan Birchall for his assistance in collecting data with SNIFS. We wish to recognize and acknowledge the very significant cultural role and reverence that the summit of Mauna Kea has always had within the indigenous Hawaiian community. This work was supported by the Director, Office of Science, Office of High Energy Physics, of the U.S. Department of Energy under Contract No. DE-AC02-05CH11231; by a grant from the Gordon \& Betty Moore Foundation; in France by support from CNRS/IN2P3, CNRS/INSU, and PNC; and in Germany by the DFG through TRR33 ``The Dark Universe." National Science Foundation Grant Number ANI-0087344 and the University of California, San Diego provided funding for HPWREN. In China support was provided from Tsinghua University 985 grant and NSFC grant No 11173017. CS was also supported by the Labex ILP (reference ANR-10-LABX-63) part of the Idex SUPER, and received financial state aid managed by the Agence Nationale de la Recherche, as part of the program ``Investissements d'avenir" under the reference ANR-11-IDEX-0004-02. This project has received funding from the European Research Council (ERC) under the European Union's Horizon 2020 research and innovation programme (grant agreement n°759194 - USNAC). The work of MVP was supported by Russian Science Foundation grant 18-72-00159. Based in part on observations obtained with the Samuel Oschin Telescope and the 60 inch Telescope at the Palomar Observatory as part of the Palomar Transient Factory project, a scientific collaboration between the California Institute of Technology, Columbia University, Las Cumbres Observatory, the Lawrence Berkeley National Laboratory, the National Energy Research Scientific Computing Center, the University of Oxford, and the Weizmann Institute of Science.

\bibliographystyle{apj}
\bibliography{proto}

\begin{thebibliography}{}
\expandafter\ifx\csname natexlab\endcsname\relax\def\natexlab#1{#1}\fi

\bibitem[{{Aldering} {et~al.}(2002){Aldering}, {Adam}, {Antilogus}, {Astier},
  {Bacon}, {Bongard}, {Bonnaud}, {Copin}, {Hardin}, {Henault}, {Howell},
  {Lemonnier}, {Levy}, {Loken}, {Nugent}, {Pain}, {Pecontal}, {Pecontal},
  {Perlmutter}, {Quimby}, {Schahmaneche}, {Smadja}, \&
  {Wood-Vasey}}]{Aldering:2002}
{Aldering}, G., {Adam}, G., {Antilogus}, P., {et~al.} 2002, in \procspie, Vol.
  4836, Survey and Other Telescope Technologies and Discoveries, ed. J.~A.
  {Tyson} \& S.~{Wolff}, 61--72

\bibitem[{{Bongard} {et~al.}(2011){Bongard}, {Soulez}, {Thi{\'e}baut}, \&
  {Pecontal}}]{Bongard:2011}
{Bongard}, S., {Soulez}, F., {Thi{\'e}baut}, {\'E}., \& {Pecontal}, {\'E}.
  2011, \mnras, 418, 258

\bibitem[{{Buton} {et~al.}(2013){Buton}, {Copin}, {Aldering}, {Antilogus},
  {Aragon}, {Bailey}, {Baltay}, {Bongard}, {Canto}, {Cellier-Holzem},
  {Childress}, {Chotard}, {Fakhouri}, {Gangler}, {Guy}, {Hsiao}, {Kerschhaggl},
  {Kowalski}, {Loken}, {Nugent}, {Paech}, {Pain}, {P{\'e}contal}, {Pereira},
  {Perlmutter}, {Rabinowitz}, {Rigault}, {Runge}, {Scalzo}, {Smadja}, {Tao},
  {Thomas}, {Weaver}, {Wu}, \& {Nearby SuperNova Factory}}]{Buton:2012}
{Buton}, C., {Copin}, Y., {Aldering}, G., {et~al.} 2013, \aap, 549, A8

\bibitem[{{Cardelli} {et~al.}(1989){Cardelli}, {Clayton}, \&
  {Mathis}}]{Cardelli:1989}
{Cardelli}, J.~A., {Clayton}, G.~C., \& {Mathis}, J.~S. 1989, \apj, 345, 245

\bibitem[{{Huang} {et~al.}(2017){Huang}, {Raha}, {Aldering}, {Antilogus},
  {Bailey}, {Baltay}, {Barbary}, {Baugh}, {Boone}, {Bongard}, {Buton}, {Chen},
  {Chotard}, {Copin}, {Fagrelius}, {Fakhouri}, {Feindt}, {Fouchez}, {Gangler},
  {Hayden}, {Hillebrandt}, {Kim}, {Kowalski}, {Leget}, {Lombardo}, {Nordin},
  {Pain}, {Pecontal}, {Pereira}, {Perlmutter}, {Rabinowitz}, {Rigault},
  {Rubin}, {Runge}, {Saunders}, {Smadja}, {Sofiatti}, {Stocker}, {Suzuki},
  {Taubenberger}, {Tao}, {Thomas}, \& {Nearby Supernova Factory}}]{Huang:2017}
{Huang}, X., {Raha}, Z., {Aldering}, G., {et~al.} 2017, \apj, 836, 157

\bibitem[{{Lantz} {et~al.}(2004){Lantz}, {Aldering}, {Antilogus}, {Bonnaud},
  {Capoani}, {Castera}, {Copin}, {Dubet}, {Gangler}, {Henault}, {Lemonnier},
  {Pain}, {Pecontal}, {Pecontal}, \& {Smadja}}]{Lantz:2004}
{Lantz}, B., {Aldering}, G., {Antilogus}, P., {et~al.} 2004, in \procspie, Vol.
  5249, Optical Design and Engineering, ed. L.~{Mazuray}, P.~J. {Rogers}, \&
  R.~{Wartmann}, 146--155

\bibitem[{{L{\'e}get} {et~al.}(2018){L{\'e}get}, {Pruzhinskaya}, {Ciulli},
  {Gangler}, {Aldering}, {Antilogus}, {Aragon}, {Bailey}, {Baltay}, {Barbary},
  {Bongard}, {Boone}, {Buton}, {Childress}, {Chotard}, {Copin}, {Dixon},
  {Fagrelius}, {Feindt}, {Fouchez}, {Gris}, {Hayden}, {Hillebrandt}, {Howell},
  {Kim}, {Kowalski}, {Kuesters}, {Lombardo}, {Lin}, {Nordin}, {Pain},
  {Pecontal}, {Pereira}, {Perlmutter}, {Rabinowitz}, {Rigault}, {Runge},
  {Rubin}, {Saunders}, {Says}, {Smadja}, {Sofiatti}, {Suzuki}, {Taubenberger},
  {Tao}, {Thomas}, \& {Nearby Supernova Factory}}]{Leget:2018}
{L{\'e}get}, P.~F., {Pruzhinskaya}, M.~V., {Ciulli}, A., {et~al.} 2018, \aap,
  615, A162

\bibitem[{{L{\'e}get} {et~al.}(2019){L{\'e}get}, {Gangler}, {Mondon},
  {Aldering}, {Antilogus}, {Aragon}, {Bailey}, {Baltay}, {Barbary}, {Bongard},
  {Boone}, {Buton}, {Chotard}, {Copin}, {Dixon}, {Fagrelius}, {Feindt},
  {Fouchez}, {Hayden}, {Hillebrandt}, {Kim}, {Kowalski}, {Kuesters},
  {Lombardo}, {Lin}, {Nordin}, {Pain}, {Pecontal}, {Pereira}, {Perlmutter},
  {Pruzhinskaya}, {Rabinowitz}, {Rigault}, {Runge}, {Rubin}, {Saunders},
  {Says}, {Smadja}, {Sofiatti}, {Suzuki}, {Taubenberger}, {Tao}, \&
  {Thomas}}]{Leget:2019}
{L{\'e}get}, P.~F., {Gangler}, E., {Mondon}, F., {et~al.} 2019, arXiv e-prints,
  arXiv:1909.11239

\bibitem[{{Nordin} {et~al.}(2018){Nordin}, {Aldering}, {Antilogus}, {Aragon},
  {Bailey}, {Baltay}, {Barbary}, {Bongard}, {Boone}, {Brinnel}, {Buton},
  {Childress}, {Chotard}, {Copin}, {Dixon}, {Fagrelius}, {Feindt}, {Fouchez},
  {Gangler}, {Hayden}, {Hillebrandt}, {Kim}, {Kowalski}, {Kuesters}, {Leget},
  {Lombardo}, {Lin}, {Pain}, {Pecontal}, {Pereira}, {Perlmutter}, {Rabinowitz},
  {Rigault}, {Runge}, {Rubin}, {Saunders}, {Smadja}, {Sofiatti}, {Suzuki},
  {Taubenberger}, {Tao}, {Thomas}, \& {Nearby Supernova Factory}}]{Nordin:2018}
{Nordin}, J., {Aldering}, G., {Antilogus}, P., {et~al.} 2018, \aap, 614, A71

\bibitem[{{Pereira} {et~al.}(2013){Pereira}, {Thomas}, {Aldering}, {Antilogus},
  {Baltay}, {Benitez-Herrera}, {Bongard}, {Buton}, {Canto}, {Cellier-Holzem},
  {Chen}, {Childress}, {Chotard}, {Copin}, {Fakhouri}, {Fink}, {Fouchez},
  {Gangler}, {Guy}, {Hillebrandt}, {Hsiao}, {Kerschhaggl}, {Kowalski},
  {Kromer}, {Nordin}, {Nugent}, {Paech}, {Pain}, {P{\'e}contal}, {Perlmutter},
  {Rabinowitz}, {Rigault}, {Runge}, {Saunders}, {Smadja}, {Tao},
  {Taubenberger}, {Tilquin}, \& {Wu}}]{Pereira:2013}
{Pereira}, R., {Thomas}, R.~C., {Aldering}, G., {et~al.} 2013, \aap, 554, A27

\bibitem[{{Rigault} {et~al.}(2018){Rigault}, {Copin}, {Aldering}, {Antilogus},
  {Aragon}, {Bailey}, {Baltay}, {Bongard}, {Buton}, {Canto}, {Cellier-Holzem},
  {Childress}, {Chotard}, {Fakhouri}, {Feindt}, {Fleury}, {Gangler},
  {Greskovic}, {Guy}, {Kim}, {Kowalski}, {Lombardo}, {Nordin}, {Nugent},
  {Pain}, {P{\'e}contal}, {Pereira}, {Perlmutter}, {Rabinowitz}, {Runge},
  {Saunders}, {Scalzo}, {Smadja}, {Tao}, {Thomas}, {Weaver}, \& {Nearby
  Supernova Factory}}]{Rigault:2018}
{Rigault}, M., {Copin}, Y., {Aldering}, G., {et~al.} 2018, \aap, 612, C1

\bibitem[{{Saunders} {et~al.}(2018){Saunders}, {Aldering}, {Antilogus},
  {Bailey}, {Baltay}, {Barbary}, {Baugh}, {Boone}, {Bongard}, {Buton}, {Chen},
  {Chotard}, {Copin}, {Dixon}, {Fagrelius}, {Fakhouri}, {Feindt}, {Fouchez},
  {Gangler}, {Hayden}, {Hillebrandt}, {Kim}, {Kowalski}, {K{\"u}sters},
  {Leget}, {Lombardo}, {Nordin}, {Pain}, {Pecontal}, {Pereira}, {Perlmutter},
  {Rabinowitz}, {Rigault}, {Rubin}, {Runge}, {Smadja}, {Sofiatti}, {Suzuki},
  {Tao}, {Taubenberger}, {Thomas}, {Vincenzi}, \& {Nearby Supernova
  Factory}}]{Saunders:2018}
{Saunders}, C., {Aldering}, G., {Antilogus}, P., {et~al.} 2018, \apj, 869, 167

\bibitem[{{Schlegel} {et~al.}(1998){Schlegel}, {Finkbeiner}, \&
  {Davis}}]{Schlegel:1998}
{Schlegel}, D.~J., {Finkbeiner}, D.~P., \& {Davis}, M. 1998, \apj, 500, 525

\end{thebibliography}
\end{document}